\documentclass[prd,superscriptaddress,twocolumn,nofootinbib,showpacs,preprintnumbers]{revtex4}

\usepackage{mathrsfs}
\usepackage{amsfonts,amssymb,amsmath}
\usepackage{graphicx,epsfig}
\usepackage{multirow}
\usepackage{verbatim}

\usepackage{color,soul}
\usepackage{xcolor}
\usepackage{framed}
\colorlet{shadecolor}{yellow}

\def\fsu5{$\cal{F}$-$SU(5)$}
\def\bfsu5{$\boldsymbol{\mathcal{F}}$-$\boldsymbol{SU(5)}$}

\def\m1half{$M_{1/2}$}
\def\m3half{$M_{3/2}$}
\def\m32{$M_{32}$}

\def\fb{${\rm fb}^{-1}$~}

\def\mt2{$M_{T2}$}
\def\x2{$\chi^2$}
\def\2b{$M_{T2}b$}

\def\bs0{$B_S^0 \rightarrow \mu^+ \mu^-$}

\begin{document}

\title{Constricted SUSY from No-Scale \bfsu5:\\ A ``Blind Spot'' for LHC Top Quark Reconstruction?}

\author{Tianjun Li}

\affiliation{State Key Laboratory of Theoretical Physics and Kavli Institute for Theoretical Physics China (KITPC),
Institute of Theoretical Physics, Chinese Academy of Sciences, Beijing 100190, P. R. China}

\affiliation{George P. and Cynthia W. Mitchell Institute for Fundamental Physics and Astronomy, Texas A$\&$M University, College Station, TX 77843, USA}

\author{James A. Maxin}

\affiliation{George P. and Cynthia W. Mitchell Institute for Fundamental Physics and Astronomy, Texas A$\&$M University, College Station, TX 77843, USA}

\affiliation{Department of Physics and Astronomy, Ball State University, Muncie, IN 47306 USA}

\author{Dimitri V. Nanopoulos}

\affiliation{George P. and Cynthia W. Mitchell Institute for Fundamental Physics and Astronomy, Texas A$\&$M University, College Station, TX 77843, USA}

\affiliation{Astroparticle Physics Group, Houston Advanced Research Center (HARC), Mitchell Campus, Woodlands, TX 77381, USA}

\affiliation{Academy of Athens, Division of Natural Sciences, 28 Panepistimiou Avenue, Athens 10679, Greece}

\author{Joel W. Walker}

\affiliation{Department of Physics, Sam Houston State University, Huntsville, TX 77341, USA}


\begin{abstract}

We describe a region of the No-Scale \fsu5 supersymmetric model with TeV-scale vector-like
matter multiplets that is potentially opaque to LHC observations based upon top quark
reconstruction.  Possessing a 100\% on-shell branching fraction for gluino to
light stop decays, the gluino--light stop mass splitting in \fsu5 lies very close to the top quark rest mass in the region
$810~ (650) \lesssim M(\widetilde{g})~ (M(\widetilde{t}_1)) \lesssim 1170~ (975)$ GeV.  For unboosted gluino pair-production,
this could render two soft top quarks in the dominant 4--top signature, greatly
reducing the likelihood for discrimination from the combinatoric background.
We therefore suggest that caution is warranted when gluino exclusion bounds on models possessing
a similar spectral character (gluino heavier than the light stop but lighter than all other squarks)
using multijet SUSY search tactics. Alternate search methods are highlighted, including
those based upon same sign dileptons and multiple heavy flavor tags, which are
more sensitive to this delicate kinematic ``blind spot'' crease. 

\end{abstract}



\preprint{ACT-5-13, MIFPA-13-20}

\maketitle


\begin{center}
\bf{\small Introduction}
\end{center}

The large hadron collider (LHC) began actively recording proton-proton collision data in 2010, amassing 
4.7 \fb at 7 TeV and 20 \fb at 8 TeV through the end of 2012. As we embark on the long
shutdown cycle and await resumption of data collection at 13-14 TeV in 2015, no conclusive signals of supersymmetry
(SUSY) have been observed thus far, with the strongest limits in multijet search simplified models for
gluino pair-production set at $M(\widetilde{g}) \gtrsim 1.35$ TeV by
ATLAS~\cite{ATLAS-CONF-2013-047} and $M(\widetilde{g}) \gtrsim 1.3$ TeV by
CMS~\cite{CMS-SUS-13-007}.  However, multijet simplified model searches generally require signatures consistent
with a gluino-mediated on-shell light stop $\widetilde{g} \to \widetilde{t}_1 t$ that is (in principle) reconstructible from
amongst the large combinatoric background, which in turn requires a relativistically substantial top quark transverse momentum $p_T$,
as might be utilized by various boosted top identification techniques~\cite{ATLAS-CONF-2012-065,CMS-PAS-JME-10-013,Kaplan:2008ie,Plehn:2010st}.
Our goal in this letter is to reiterate that if nature makes use of a locally degenerate stop-gluino SUSY mass structure,
where the event topologies probed by ATLAS and CMS are constricted
(compare to ``compressed'' SUSY~\cite{LeCompte:2011cn,LeCompte:2011fh} and ``stealth'' SUSY~\cite{Fan:2012jf})
by the narrow phase space to produce off-shell or low $p_T$ top quarks from gluinos,
and only the decay mode $\widetilde{g} \to \widetilde{t}_1 t$ is kinematically allowed,
then the baseline advertisements for gluino and stop mass limits established from searches that rely on this
methodology may be inapplicable or overly strong.

\begin{center}
\bf{\small The \bfsu5 Model}
\end{center}

A concrete SUSY model fitting the prior description is No-Scale \fsu5, ({\it cf.}
Refs.~\cite{Li:2010ws,Li:2010mi,Li:2010uu,Maxin:2011hy,Li:2011xu,Li:2011in,Li:2011ab,Li:2012yd,Li:2013hpa,Li:2013naa}
and references therein), which is based upon the tripodal foundations of
the dynamically established boundary conditions of No-Scale Supergravity, the Flipped $SU(5)$ Grand Unified Theory (GUT),
and hypothetical TeV-scale ``{\it flippon}'' vector-like super-multiplets motivated within local
F-theory model building. The union of these features has been demonstrated to naturally solve many
standing theoretical difficulties, and to positively compare with real world experimental observation.

Written in full, the gauge group of Flipped $SU(5)$ is $SU(5)\times U(1)_{X}$, which can be embedded into $SO(10)$.
The generator $U(1)_{Y'}$ is defined for fundamental five-plets as $-1/3$ for the triplet members, and $+1/2$ for the doublet.
The hypercharge is given by $Q_{Y}=( Q_{X}-Q_{Y'})/5$. There are three families of Standard Model (SM)
fermions, whose quantum numbers under the $SU(5)\times U(1)_{X}$ gauge group are
\begin{equation}
F_i={\mathbf{(10, 1)}} \quad;\quad {\bar f}_i={\mathbf{(\bar 5, -3)}} \quad;\quad {\bar
l}_i={\mathbf{(1, 5)}},
\label{eq:smfermions}
\end{equation}
where $i=1, 2, 3$. There is a pair of ten-plet Higgs for breaking the GUT symmetry, and a pair
of five-plet Higgs for electroweak symmetry breaking (EWSB).
\begin{eqnarray}
& H={\mathbf{(10, 1)}}\quad;\quad~{\overline{H}}={\mathbf{({\overline{10}}, -1)}} & \nonumber \\
& h={\mathbf{(5, -2)}}\quad;\quad~{\overline h}={\mathbf{({\bar {5}}, 2)}} &
\label{eq:Higgs}
\end{eqnarray}

SUSY must be broken around the TeV scale, as parameterized in minimal supergravities (mSUGRA)
by universal SUSY-breaking ``soft terms'' which include the gaugino mass $M_{1/2}$, scalar mass $M_0$ and the trilinear coupling $A$.
The ratio of the low energy Higgs vacuum expectation values (VEVs) $\tan \beta$, and the sign of
the SUSY-preserving Higgs bilinear mass term $\mu$ are also undetermined, while the magnitude of the $\mu$ term and its bilinear soft term $B_{\mu}$
are determined by the $Z$-boson mass $M_Z$ and $\tan \beta$ after EWSB.  In the simplest No-Scale scenario,
$M_0$=A=$B_{\mu}$=0 at the unification boundary, while the complete collection of low energy SUSY breaking soft-terms evolve down 
single non-zero parameter $M_{1/2}$.  Consequently, the particle spectrum will be proportional to $M_{1/2}$ at leading order,
rendering the bulk ``internal'' physical properties invariant under an overall rescaling. The matching condition between the low-energy value of
$B_\mu$ that is demanded by EWSB and the high-energy $B_\mu = 0$ boundary is notoriously difficult to reconcile under the
RGE running. The present solution relies on modifications to the $\beta$-function coefficients that are generated by the {\it flippon} loops.

Crucially, application of the non-trivial boundary condition $B_\mu =0$ appears to come into its
own only when applied at a unification scale approaching the Planck mass $M_{\rm Pl}$~\cite{Ellis:2001kg,Ellis:2010jb,Li:2010ws}.
There is an intriguing possibility in the flipped $SU(5)$ GUT that the natural decoupling of an intermediate unification for the
$SU(2)_{\rm L} \times SU(3)_{\rm c} \Rightarrow SU(5)$ subgroup from a final unification
with the remixed hypercharge $U(1)_{\rm X}$ might be exploited to push the upper unification within
the targeted proximity of $M_{\rm Pl}$.  With only the field content of the minimal supersymmetric
standard model (MSSM), the second phase of running in the renormalization group equations (RGEs) is quite short.
There are two explicitly realizable vector-like multiplet configurations that may be 
introduced around the TeV scale while avoiding a strong Landau pole~\cite{Jiang:2006hf},
each of which enhance the (formerly negative) one-loop $\beta$-function coefficient of the strong coupling,
such that it becomes precisely zero. The flatness in the running of the strong coupling $\alpha_{\rm s}$ creates a wide
gap between the couplings $\alpha_{32} \simeq \alpha_{\rm s}$ and the much smaller $\alpha_{\rm X}$
at the intermediate unification. This gap can only be closed by a very significant secondary running
phase, which may thus elevate the final unification scale by the necessary 2-3 orders of magnitude~\cite{Li:2010dp}.

The effect of these changes to the $\beta$-function coefficients on the gluino is direct in the running down
from the high energy boundary, leading to the relation $M_3/M_{1/2} \simeq \alpha_3(M_{\rm Z})/\alpha_3(M_{32}) \simeq \mathcal{O}\,(1)$
and precipitating a conspicuously light gluino mass assignment. Likewise, the large mass splitting expected from the heaviness
of the top quark, via its strong coupling to the Higgs, is responsible for a rather light stop squark $\widetilde{t}_1$.
The distinctively predictive $M({\widetilde{t_1}}) < M({\widetilde{g}}) < M({\widetilde{q}})$ mass hierarchy of a light stop
and gluino, both much lighter than all other squarks, is stable across the full No-Scale \fsu5 model space.  In conjunction
with uniform mass hierarchy dilation in proportion to $M_{1/2}$, it follows that this model will experience a phase transition
from off-shell to on-shell decays of the gluino to a top quark and stop squark pair, as the mass spacing relaxes 
at heavier global scales.  It is this property, and particularly the kinematic constriction occurring just
subsequent to that ``blind spot'' transition, which compels the associated top quark production to
be relatively kinematically soft, and thus exceedingly vulnerable to failed reconstruction.  This point
drives the discussion of the present letter, although the conclusions are equally applicable to
any model with a similar spectral character.

\begin{figure}[htp]
        \centering
        \includegraphics[width=0.5\textwidth]{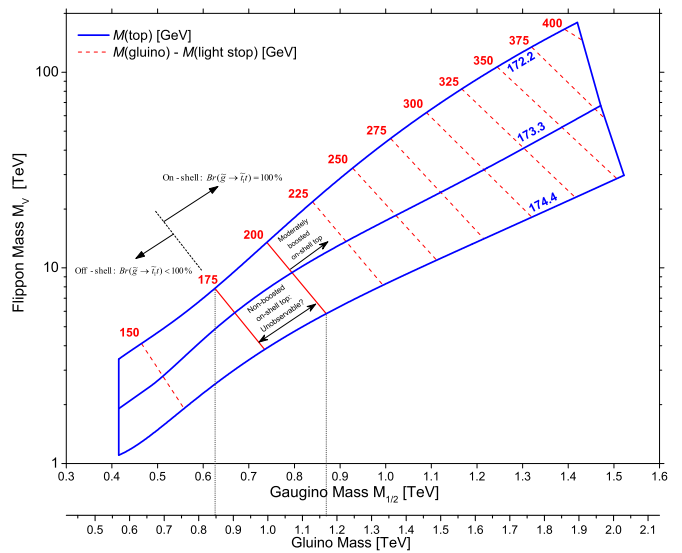}
        \caption{Constrained model space of No-Scale \fsu5 as a function of the gaugino mass $M_{1/2}$,
gluino mass $M(\widetilde{g})$, and flippon mass $M_V$. The thick lines demarcate the upper and lower exterior
boundaries as defined by a top quark mass of $m_t = 173.3 \pm1.1$ GeV. The left edge is marked by the LEP
constraints, while the right edge depicts where the Planck relic density can no longer be maintained due to an
LSP and light stau mass difference less than the on-shell tau mass. All model space within these boundaries
satisfy the Planck relic density constraint $\Omega h^2 = 0.1199 \pm 0.0027$ and the No-Scale requirement $B_{\mu}=0$.
The dashed and solid lines within the interior denote the $M(\widetilde{g}) - M(\widetilde{t}_1)$ mass difference,
with the off-shell to on-shell transition labeled. We make special note of the
$175 \lesssim M(\widetilde{g}) - M(\widetilde{t}_1) \lesssim 200$ GeV region of unboosted top quarks in
the decay mode $\widetilde{g} \to \widetilde{t}_1 t$ that could escape distinction from the combinatoric background at the LHC.}
        \label{fig:wedge}
\end{figure}

In order to make specific quantitative statements regarding the \fsu5 model, including the detailed SUSY particle mass structure,
we employ an industry standard suite of tools, including {\tt MicrOMEGAs 2.1}~\cite{Belanger:2008sj} and {\tt SuSpect 2.34}~\cite{Djouadi:2002ze},
where proprietary modifications have been made to incorporate the flippon RGEs.
We find that $M(\widetilde{g}) \simeq m_t + M(\widetilde{t}_1)$, immediately triggering onset of a unity (100\%) branching fraction
for the associated on-shell decay, above a threshold mass scale of $M(\widetilde{g}) \sim 810$ GeV ($M(\widetilde{g}) \sim 980$)
for a top quark mass of $m_t = 172.2$ GeV ($m_t = 174.4$ GeV) ~\cite{Li:2013naa}, whereas the gluino-mediated light
stops are off-shell in \fsu5 at lighter scales.
The full extent of the \fsu5 ``bare-minimally'' phenomenologically constrained parameter space~\cite{Li:2013naa} (prior
to application of LHC collider limits), stretches from a lower gluino mass around $M(\widetilde{g}) \simeq$ 540 GeV
(corresponding to an $M_{1/2} \simeq$ 385 GeV) to an upper termination around $M(\widetilde{g}) \simeq$ 2.0 TeV
(corresponding to an $M_{1/2} \simeq$ 1.5 TeV), as depicted in Figure (\ref{fig:wedge}). 

The described spectrum generates a unique event topology starting from the pair production of heavy squarks
$\widetilde{q} \widetilde{\overline{q}}$, except for the light stop, in the initial hard scattering process,
with each squark likely to yield a quark-gluino pair $\widetilde{q} \rightarrow q \widetilde{g}$.  Each gluino may be expected
to produce events with a high multiplicity of virtual stops, via the (possibly off-shell) $\widetilde{g} \rightarrow \widetilde{t}t$
transition, which in turn may terminate into hard scattering products such as $\rightarrow W^{+}W^{-} b \overline{b} \widetilde{\chi}_1^{0}$
and $W^{-} b \overline{b} \tau^{+} \nu_{\tau} \widetilde{\chi}_1^{0}$, where the $W$ bosons will produce mostly hadronic jets and some leptons.
The final state products may then consistently exhibit a net content of eight or more hard jets emergent from a single squark pair production event,
passing through a single intermediate gluino pair, resulting after fragmentation in a spectacular signal of ultra-high multiplicity final state jet events.

The preference of the \fsu5 model for ultra-high multi-jet multiplicities likewise selects searches that
probe correspondingly high jet count signatures, {\it e.g.}~Ref~\cite{ATLAS-CONF-2013-047,ATLAS-CONF-2013-054,ATLAS-CONF-2013-061,CMS-SUS-13-007},
as ideal metrics for comparison against Monte-Carlo simulation of this model.
However, many searches of this variety (we shall focus in this letter on the example of Ref.~\cite{ATLAS-CONF-2013-054})
are deeply susceptible to the described intrinsic kinematic limitation on attempts to reconstruct the boosted top quark decay.
It is certainly possible for alternative search strategies that do not similarly rely on top quark
reconstruction to provide cross-coverage of the parameter space afflicted by this shortcoming,
such that the union of exclusion perimeters is continuous, and we shall close the current letter by
pointing out likely candidates for this purpose (focusing on the examples of Refs.~\cite{ATLAS-CONF-2013-061,ATLAS-CONF-2013-007,Chatrchyan:2012paa})
with respect to the \fsu5 model; nevertheless, our current interest is to illuminate the features of this interesting
kinematic dead zone, within the context of the present concrete example.

\begin{center}
\bf{\small The Top Quark ``Blind Spot''}
\end{center}

Highly boosted top quarks were not produced at the Tevatron, though they are being produced in copious
amounts at the LHC. Considering that the on-shell $\widetilde{g} \to \widetilde{t}_1 t$ SUSY process is
expected to generate highly boosted top quarks, much effort has been directed at boosted top quark
reconstruction methods for the LHC (for example, see
Refs.~\cite{Kaplan:2008ie,Plehn:2009rk,Plehn:2010st,Abdesselam:2010pt,Berger:2011af}). The
top quark reconstruction algorithm {\tt HEPTopTagger}~\cite{Plehn:2009rk,Plehn:2010st}, which is
based upon the Cambridge/Aachen
algorithm~\cite{Dokshitzer:1997in,Wobisch:1998wt,Cacciari:2005hq}, requires that the combined
$p_T$ of the three subjets in $t \to bW \to bjj$ must exceed 200 GeV.  For boosted tops, the $bjj$ are
anticipated to be highly collimated within a small $\Delta R_{bjj}$ for $p_T \gtrsim 200$ GeV.
This boosted top reconstruction threshold establishes a corresponding lower
limit on gluino--light stop mass splitting in the decay channel $\widetilde{g} \to \widetilde{t}_1 t$ of
$M(\widetilde{g}) - M(\widetilde{t}_1) \gtrsim 200$ GeV for unboosted gluino pair-production.
If this threshold is not met there is a substantial likelihood that the top quark will
be too soft for detection when the parent gluinos are produced in unboosted pairs.
For an attempt to preserve unboosted top quarks in SUSY final states, see, for example, Ref.~\cite{Dutta:2011gs}.

\begin{figure}[htp]
        \centering
        \includegraphics[width=0.5\textwidth]{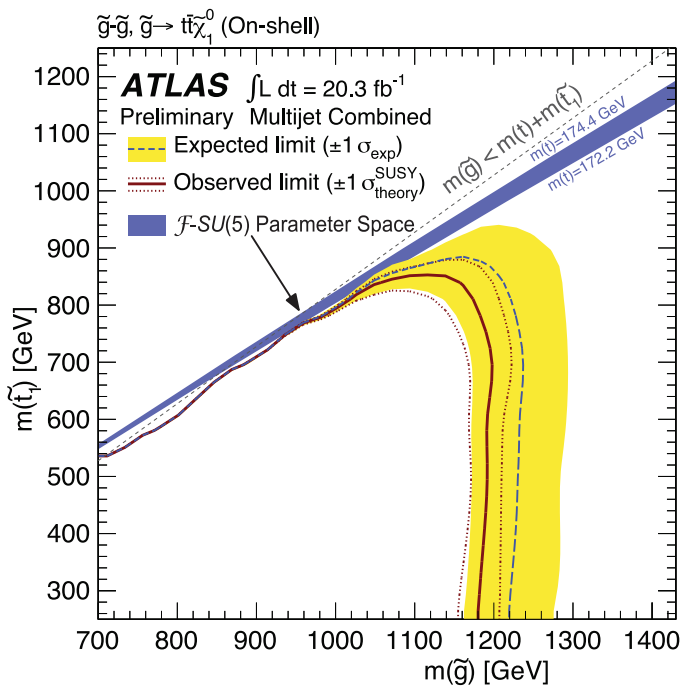}
        \caption{The No-Scale \fsu5 parameter space of Figure (\ref{fig:wedge}) is overlaid on Figure (9) of
the ATLAS large jet multiplicity SUSY search of Ref.~\cite{ATLAS-CONF-2013-054}. The solid streak represents
the \fsu5 model space, which lives in the unprobed crease between the on-shell boundary and the ATLAS
observable limit for the gluino--light stop multijet simplified model depicted here. The ATLAS limits shown
here are for a fixed LSP mass of 60 GeV, and will thus shift by a small amount for a wider range of LSP masses.}
        \label{fig:atlas}
\end{figure}

The off-shell to on-shell transition of the \fsu5 model therefore poses a particular dilemma for gluino masses from
$M(\widetilde{g}) \sim 810 - 1170$ GeV, as the on-shell gluino-mediated light stops in \fsu5 could escape
detection at the LHC due to the gluino-light stop mass splitting residing very close to the top quark rest mass.
The region of concern is distinguished from the bulk
\fsu5 model space~\cite{Li:2013naa} in Figure (\ref{fig:wedge}).  This potential ``blind spot''
is further visually exhibited in Figure (\ref{fig:atlas}), where we overlay the No-Scale \fsu5 parameter space on
top of Figure (9) from the ATLAS large jet multiplicity SUSY search of Ref.~\cite{ATLAS-CONF-2013-054}.
Curiously, the \fsu5 model space lives largely within the small unprobed crease between the dashed
on-shell line and the thick solid line representing the ATLAS observable lower bound on the gluino versus
the light stop mass.  Note that since the ATLAS simplified model plot adopted for use in Figure
(\ref{fig:atlas}) is for a fixed lightest supersymmetric particle (LSP) mass of 60 GeV, the observed
limits will shift a small amount for a wider application of LSP masses, although we only use it here as a
general aid to illustrate the relationship between the gluino--light stop mass splitting in \fsu5 and the
unprobed region of unboosted top quarks from gluino decays for a particular LHC search.

Once the gluino-mediated light stops transition to on-shell in \fsu5, the branching fraction for
$\widetilde{g} \to \widetilde{t}_1 t$ jumps to 100\%~\cite{Li:2013naa}. Therefore, gluino
pair-production will dominantly result in a 4--top signal $\widetilde{g} \widetilde{g} \to t
\overline{t} t \overline{t} + 2 \widetilde{\chi}_1^0$, distinguishable from the Standard Model
background by the corresponding large missing energy component, with an associated large multijet
signal serving as the characteristic collider signature~\cite{Maxin:2011hy}. On the other
hand, if two of these top quarks possess $p_T \lesssim 200$ GeV, then the odds of discriminating the
residual portion of the event from the combinatoric background decrease substantially.
If, in fact, the physical gluino mass were $810 \lesssim M(\widetilde{g}) \lesssim 1170$ GeV, then
a variety of other search strategies more sensitive to the $M(\widetilde{g}) \sim M(\widetilde{t}_1 + m_t)$ crease 
would be essential in order to make that discovery.  Caution is therefore warranted in conclusively
excluding these light gluinos; complementary search channels that are potentially suitable to this
purpose will be described as an afterword to the present letter.

The predicament presented here solely affects gluino-mediated channels, as the top quark in $\widetilde{g} \to \widetilde{t}_1 t$ could be unboosted,
and furthermore, the branching fraction for $\widetilde{g} \to q \overline{q} + \widetilde{\chi}_1^0$
is 0\% once the gluino-mediated light stop decay goes on-shell at about $M(\widetilde{g}) \gtrsim 810$
GeV. Consequently, light stop pair-production in the context of \fsu5 remains fully observable. The
branching fraction for the decay mode $\widetilde{t}_1 \to t + \widetilde{\chi}_1^0$ in this ``blind
spot'' region runs from 60 -- 75\%~\cite{Li:2013naa}, though the LHC Collaboration's simplified model
limits for light-stop pair-production searches typically assume a 100\% branching fraction for light
stop decays to the top quark or chargino. Nevertheless, the lower bounds established on the light stop mass
from the 20 \fb data by ATLAS~\cite{ATLAS-CONF-2013-024,ATLAS-CONF-2013-037} at
$M(\widetilde{t}_1) \gtrsim 620$ GeV and CMS~\cite{CMS-SUS-13-011} at $M(\widetilde{t}_1) \gtrsim
650$ GeV can be reasonably applied to \fsu5 without additional concern. Due to the linear relationship between
the gluino and light stop in \fsu5, a limit on the light stop mass can be directly correlated to a corresponding bound on the
gluino mass, which translates to approximately $M(\widetilde{g}) \gtrsim 810$ GeV in this case, right at
the lower threshold of the potential gluino ``blind spot'' in \fsu5.

In addition to light stop pair-production, the observed limits from squark pair-production with
decoupled gluinos via $\widetilde{q} \widetilde{q} \to q \overline{q} + 2 \widetilde{\chi}_1^0$ dijet
events remain unaffected in this region.
For squark pair-produced dijet events, both the ATLAS observed squark limit of $M(\widetilde{q})
\gtrsim 740$ GeV~\cite{ATLAS-CONF-2013-047} (corresponding to $M(\widetilde{g}) \gtrsim 500$ GeV in
\fsu5) and the CMS observed squark limit of $M(\widetilde{q}) \gtrsim 800$
GeV~\cite{Chatrchyan:2013lya} are considerably weaker than the light-stop pair-production limits.

We emphasize that the present analysis has been conducted in the spirit of the simplified model adopted by
the ATLAS collaboration in Figure (\ref{fig:atlas}), wherein the initial hard scattering is confined
solely to gluino pair production. In particular, in a realistic model with the mass hierarchy of No-Scale
\fsu5, there may also be contributions to the same final states from initial hard scattering into
squark--squark or squark--gluino pairs. Although these production modes are substantially
subdominant in \fsu5 by cross-section, they may potentially be elevated into a majority population of
events surviving the application of selection cuts, due firstly to the extra jet(s) afforded by squark to
gluino decay $\widetilde{q} \to q \widetilde{g}$, and secondly to the enhanced likelihood that the
resulting gluino may inherit a substantial Lorentz boost. Nevertheless, deep suppression of the primary
intended discovery target argues against the suitability of search channels devised for gluino pair
production discovery as strong expected limit discriminants in models adjacent to \fsu5, within the
described kinematic ``blind spot''.

As an aside, limits attributable to squark--squark and squark--gluino pair production
may potentially be softened by an independent mechanism, if modifications to the SUSY breaking soft terms
are allowed that include relatively large scalar masses for the first two generation sfermions. In this
scenario, the original squark--squark and squark--gluino pair production events can be highly
suppressed while preserving the described impact of the No-Scale \fsu5 ``blind spot'', since the
electroweak gauge symmetry breaking is dominantly related to the Higgs sector and the third family of
Standard Model fermions.
One circumstance in particular that would call for such deviations to the SUSY breaking soft terms to be
investigated is if light stop pair-production searches indicate the presence of light stops in the region
$650 \lesssim M(\widetilde{t}_1) \lesssim 975$ GeV (which would correlate to $810 \lesssim
M(\widetilde{g}) \lesssim 1170$ GeV), yet the $\widetilde{q} \widetilde{q} \to qq + \widetilde{g}
\widetilde{g}$ and $\widetilde{q} \widetilde{g} \to q + \widetilde{g} \widetilde{g}$
pair-production channels offer no evidence of gluinos and light stops in the LHC data when using the
anticipated first and second generation squark masses in \fsu5~\cite{Li:2013naa}.

In conclusion, we remarked in this letter that the No-Scale \fsu5 model lives partially within a
particularly difficult region of phase space for the LHC to probe, {\it i.e.} a ``blind spot''.
Due to this circumstance,  prudence is advisable with regards to the setting of exclusion bounds
on gluino masses within the region $810 \lesssim M(\widetilde{g}) \lesssim 1170$ GeV, which correlates to light
stop masses of $650 \lesssim M(\widetilde{t}_1) \lesssim 975$ GeV and LSP masses of $120 \lesssim
M(\widetilde{\chi}_1^0) \lesssim 180$ GeV.  In order to study this \fsu5 mass interval, 
complementary search strategies are necessary, which are invulnerable to the kinematic constriction
of daughter top quark particles, including searches targeting direct light stop pair production.
The forthcoming 13-14 TeV LHC will be required in order to examine the \fsu5 model space
extending beyond this problematic ``blind spot'' region.


\begin{center}
\bf{\small Afterword}
\end{center}

Shortly after publishing the first version of this note, an interesting study was released
by the ATLAS collaboration~\cite{ATLAS-CONF-2013-061} that elegantly complements the search
space probed by the primary experimental case example~\cite{ATLAS-CONF-2013-054} considered
here.  In this work, zero and one lepton signal regions are statistically combined to offset
the described limitations on sensitivity to final states profused by top quarks, supplemented
by additional selections that include a triple heavy flavor tag requirement, a large effective
event mass, and 4 to 7 total jets.  In particular,
the Gluino-Stop I and II models are applicable to the \fsu5 scenario. Both cases assume
100\% branching for the (on-shell) $\widetilde{g} \rightarrow \widetilde{t}_1 t$ decay, and specifically adopt the
\fsu5 spectral ordering $M({\widetilde{t_1}}) < M({\widetilde{g}}) < M({\widetilde{q}})$. In the
type I scenario, the stop subsequently decays exclusively to a bottom quark and chargino pair
$\widetilde{t}_1 \rightarrow b \widetilde{\chi}_1^\pm$,
while the type II scenario exclusively specifies a top quark and neutralino product
$\widetilde{t}_1 \rightarrow t \widetilde{\chi}_1^0$.
The quoted exclusion bounds for each
of these two simplified models, respectively, are (850,1150)~GeV for the light stop
$\widetilde{t}_1$, and (1050,1320)~GeV for the gluino $\widetilde{g}$.
The \fsu5 model is sensitive to production in both of these modes.

The ATLAS study described in Ref.~\cite{ATLAS-CONF-2013-007} sidesteps
requirements on large missing energy while probing gluino decays into a light stop
by requiring a same-sign dilepton in association with multiple heavy flavor tags.  Limits are again provided for the simplified model categories most relevant to \fsu5, but the bounds in this case are
substantially softer, in the vicinity of 1~TeV for the gluino in both stop squark decay modes.

A CMS search conducted with a very similar spirit~\cite{Chatrchyan:2012paa}, but using
only half of the currently available dataset, likewise sets a gluino bound around 1~TeV
for the most relevant (A2) production channel.

In order to better interpret the rather stringent limits suggested by Ref.~\cite{ATLAS-CONF-2013-061}, 
an independent Monte-Carlo analysis~\cite{Stelzer:1994ta,Alwall:2007st,Sjostrand:2006za,PGS4},
with replication of the specified selection cuts~\cite{Walker:2012vf}, has been undertaken.
Our results indeed suggest that regions of parameter space below $M_{1/2} \sim 950-990$~GeV, corresponding at the upper end to roughly 1100~GeV and 1300~GeV for the stop and gluino, are disfavored in \fsu5. It is emphasized that this development does not exclude the featured model, whose viable
parameter space extends somewhat beyond $M_{1/2} \sim 1.5$~TeV~\cite{Li:2013naa}. Neither, does it negate the basic observation that searches dependent upon top quark reconstruction tagging will be relatively desensitized for models in the spectral vicinity of No-Scale \fsu5. On the contrary, it strongly validates the essential role of complementary event analyses in patching over the limitations intrinsic to various distinct methodologies.


\begin{center}
\bf{\small Acknowledgments}
\end{center}

We thank Teruki Kamon for enlightening discussions on top quark reconstruction methodology.
This research was supported in part by the DOE grant DE-FG03-95-Er-40917 (DVN), the Natural Science
Foundation of China under grant numbers 10821504, 11075194, 11135003, and 11275246 (TL),
the Mitchell-Heep Chair in High Energy Physics (JAM), and the Sam Houston State University 2013 Faculty
Research Grant program (JWW).  We also thank Sam Houston State University for providing
high performance computing resources.


\bibliography{bibliography}

\end{document}